\documentclass[conference]{IEEEtran}

\usepackage{cite}
\usepackage{amsmath,amssymb,amsfonts}
\usepackage{algorithm}
\usepackage[noend]{algorithmic}
\usepackage{graphicx}
\usepackage{textcomp}
\usepackage{xcolor}
\usepackage{makecell}
\usepackage{hyperref}

\def\BibTeX{{\rm B\kern-.05em{\sc i\kern-.025em b}\kern-.08em
    T\kern-.1667em\lower.7ex\hbox{E}\kern-.125emX}}
\begin{document}

\title{Automatic Storage Structure Selection for hybrid Workload}

\author{\IEEEauthorblockN{Hongzhi Wang, Yan Wei and Hao Yan}
\IEEEauthorblockA{\textit{School of Computer Science and Technology} \\
\textit{Harbin Institute of Technology}\\
Harbin, China \\
\{wangzh, 18s003043\}@hit.edu.cn, 949514033@qq.com}
}

\maketitle

\begin{abstract}
In the use of database systems, the design of the storage engine and data model directly affects the performance of the database when performing queries. Therefore, the users of the database need to select the storage engine and design data model according to the workload encountered. However, in a hybrid workload, the query set of the database is dynamically changing, and the design of its optimal storage structure is also changing. Motivated by this, we propose an automatic storage structure selection system based on learning cost, which is used to dynamically select the optimal storage structure of the database under hybrid workloads. In the system, we introduce a machine learning method to build a cost model for the storage engine, and a column-oriented data layout generation algorithm. Experimental results show that the proposed system can choose the optimal combination of storage engine and data model according to the current workload, which greatly improves the performance of the default storage structure. And the system is designed to be compatible with different storage engines for easy use in practical applications.
\end{abstract}


\section{Introduction}
Large companies, especially those in the Internet industry, are keen to collect and analyze data. After a large amount of data is generated in trades or mobile applications, they are quickly used for decision support, business intelligence, and personalized recommendation. In this process, the data needs to be capable of processing under both transactional and analytical workloads, which is also known as hybrid transaction-analytical processing (HTAP). Data engineers need to balance hardware costs and performance, and provide the HTAP systems suitable for their business scenarios. Data engineers need to investigate and compare databases from different vendors and select a solution within a limited time. Such a solution often involves a combination of multiple databases that support different workload types, which may include an OLAP database and an OLTP database (or NOSQL that supports write-heavy workloads), and is connected and synchronized through ETL tools\cite{c13}.

The development of the database has entered the stage of designing for specific tasks, and different databases have different designs and advantages. As we know, "One size not fit all\cite{c11}", the database chooses different models from the design of the underlying storage engine in order to adapt to the needs of different workloads. These databases are different from each other and cannot replace each other\cite{c12}. For example, even databases based on the same LSM model have different types to weigh the performance of write and scan operations\cite{c6}. Among the rich choices, it is difficult for data engineers to choose the most suitable type based on the database description or experience. Even more, unfortunately, the workloads created by the business are not static, they may shift over time. The selected database may no longer be suitable for new workloads. In summary, data engineers need to face the following issues when building HTAP systems:

\begin{itemize}
\item To combine different databases, engineers may need to use ETL tools to synchronize in different databases. The synchronization process generally pulls the latest data from the OLTP database and imports it into the OLAP database. In this case, analytical queries cannot directly access the latest data in the OLTP database.
\item Data engineers need to understand the principles of different databases, performance evaluation, and even experience to select the database type in the system. Considering the capabilities and experience of data engineers, this may take several days or even months to investigate and experiment.
\item The workload may change dynamically with time. Such changes may result in the original design no longer being optimal, or even failing to meet performance requirements.
\end{itemize}

In recent years, self-driven databases have liberated data engineers from tedious tasks\cite{speedup}, while these databases prefer to be developed as a new system. They have emerged that can dynamically adjust the storage design to achieve high-performance processing of HATP workloads. However, such systems require expensive hardware (such as in-memory database using large amounts of memory\cite{c14} or global shared memory\cite{c15}). Moreover, the immaturity of these systems also prevents data engineers from choosing them.

For enterprises, the best solution still uses the existing mature database for system design. Thus, instead of establishing a new system, we attempt to take the advantages of the existing database to solve the problem of automatic design of the HTAP system, which can make full use of the community and low-cost features of these databases.

Motivated by this, in this paper, we develop the automatic design method for the HTAP database based on existing databases. The focus is to select proper storage structure according to the features of hybrid workload. To achieve this goal, we requires an effective way to evaluate the cost of the workload execution on the data storage. With the consideration of the diversity of data storage structures, we develop a cost evaluation approach based on machine learning without the knowledge of database internals. Such that the storage structure is selected according to such estimated cost. The contributions of this paper are summarized as follows:

\begin{itemize}
\item In this paper, we regard the data layout under a specific storage engine as a storage structure and proposes a storage structure selection system based on learning cost. The proposed system can automatically perform storage structure generation, evaluation, and conversion.

\item We propose a learning method for the storage engine cost model, which does not require the internal knowledge of the storage engine.

\item A novel benchmark is proposed for the establishment of learning models, which can dynamically generate data schemas and workloads for performance data collection.

\item To increase flexibility, this paper proposes a column-oriented data layout recommendation algorithm, which is used to generate storage structure candidates.
\end{itemize}

The structure of this paper is as follows. Section II introduces the architecture of the system, including the explanation of each key module and the workflows of the system function. Then in the Section III, we introduce the method of building a storage engine learning model, including how to design features, how to collect training data, and how to build a more accurate model. In the Section IV, we introduce the selection method of storage structure, especially the column-oriented data layout recommend algorithm. Section V conducts experimental evaluation of the system. Finally, summary and discussion of the future work in Section VI.

\section{System overview}
In this section, we propose the architecture of the storage structure recommendation system. First, we summarize the methodology for relevant systems in Section II-A, which will help to understand the proposed system. Then we introduced the main modules of the system in Section II-B, especially the candidate execution module. Finally, we introduced the workflows related to storage structure recommendations in Section II-C.
\subsection{Methodology}

The purpose of the self-driving database is to establish a kind of self-designing and self-tuning database management system. The optimization task of automatically selecting the storage structure plays an important role in self-driving databases. During the decades of autonomous database development, a multi-level theoretical system has been established \cite{c1}, which acts on the knobs tuning, algorithm picking, and even the self-design of database components. The current research on the search method for algorithm picking problems \cite{c2} can be simplified into the following three steps:

\begin{itemize}
    \item \textbf{Pruning:}The original search space for algorithm picking problems may be huge, so it’s better to prune the search space of the problem by some methods. For example, the multi-index optimization problem can be transformed into a combination of single-index optimization problems, and the queries that have little effect can be removed in the workload to narrow the input.

    \item  \textbf{Candidates Generation:}
    Traverse the search space, and generate candidates by enumerating solutions using related rules. If the original problem is disassembled into several sub-problems, the candidates of the original problem need to be generated according to the candidates of the sub-problems.

    \item \textbf{Evaluation:}Choose an optimal solution among the candidates by calculating costs. For the candidate cost that is easy to be tested, we can run a set of experiments to find the best candidate. For candidates that cannot be easily tested, the What-If cost model needs to be established\cite{c16}.
\end{itemize}

\begin{figure*}[!t]
    \centering
    \includegraphics[width=5in]{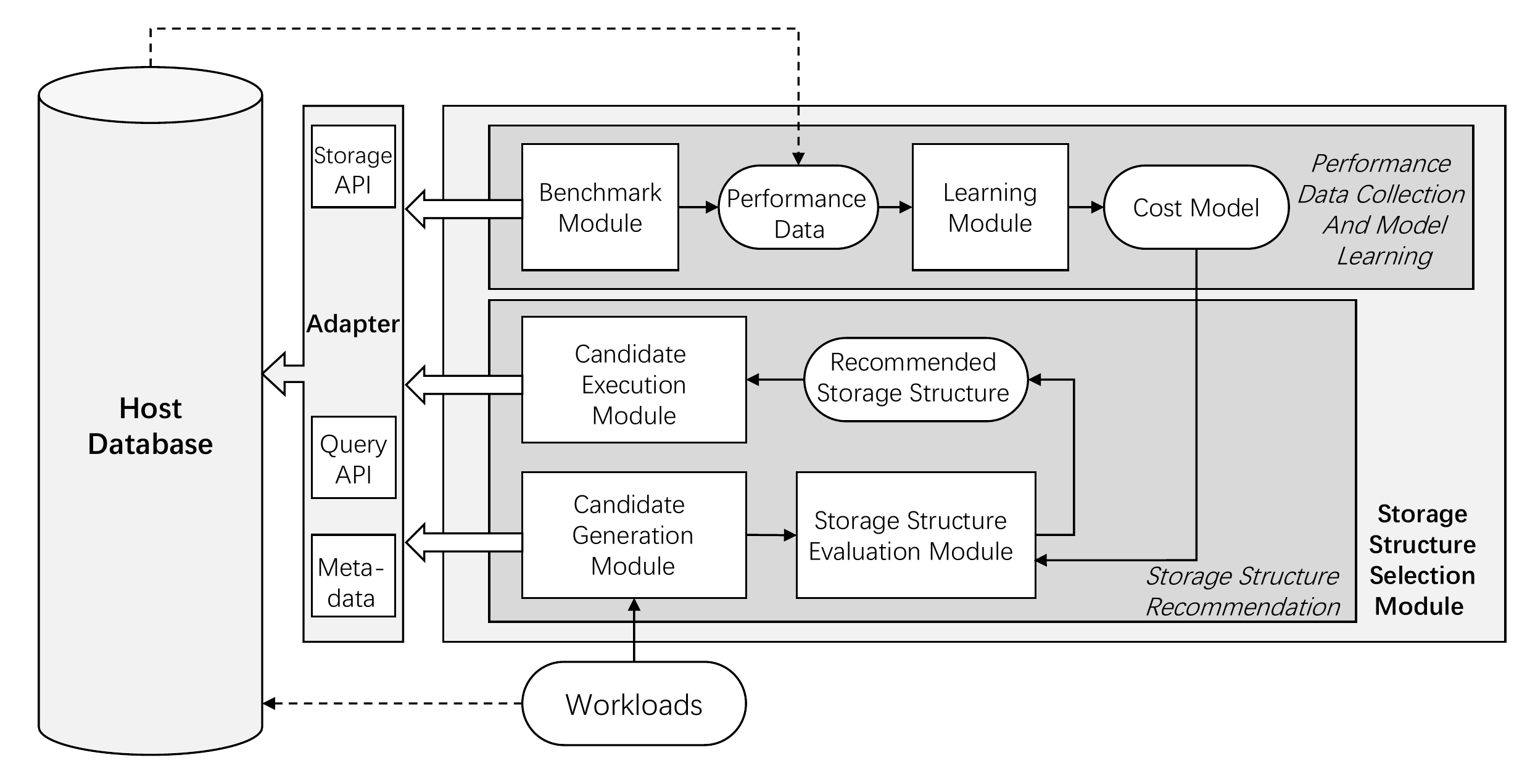}
    \caption{System architecture}
\label{figure1}
\end{figure*}

The system proposed in this paper follows the search steps mentioned above. In order to provide compatibility with existing storage engines, our system is designed with low coupling to the host database. The system is divided into three parts as shown in
Fig. \ref{figure1}, namely the host database, the database adapter, and the storage structure selection module. The workload is analyzed in the storage structure selection module to obtain the candidate storage structures, and the optimal storage structure is selected by comparing the cost of the candidates. Since the storage structure scheme has a relatively high cost to be tested, it is not feasible to run experiments to see which one is the best choice, in which case we propose a series of What-If cost models to achieve the goal of candidate evaluation.

\subsection{Storage structure selection module}

The storage structure selection module is divided into five submodules, namely benchmark module, learning module, candidate generation module, storage structure evaluation module and candidate execution module. The submodules are introduced as follows:

\begin{itemize}
    \item
    \textbf{Benchmark module:} This module manages the execution and performance data collection of the benchmark. The benchmark is a simulation on the host database, based on generated workload with generated data. It can be used to obtain the performance data of the host database under different storage engines and different operations. The benchmark is customized for the proposed system and will be discussed in Section III-B.
    \item
    \textbf{Learning module:}  This module learns and updates the cost model. The cost model is learned by benchmarking result data and the performance data collected at runtime, and is used for the cost estimation of different storage structures under a given workload. The learning model establish method will be discussed in Section III-C.
    \item
    \textbf{Candidate generation module:} This module analyzes the recent workload and then generates a set of candidates which are the data layout and storage engine combinations. This module takes workloads and table meta data as input, and workloads are pruned before being used for candidate generation. The details of the data layout recommendation algorithm is discussed in Section IV.
    \item
    \textbf{Storage structure evaluation module:} This module enumerates the cost of each candidate storage structure, and obtains the optimal storage structure based on the comparison. Since the workload needs to be converted into a set of query plans before it can be used for cost estimation, this module requires a connector to the built-in query planner in the host database.
    \item
    \textbf{Candidate execution module:} This module applies the optimal storage structure to the target data partition if required. Whether it is the switching of the storage engine or the transformation of the data layout, it is necessary to convert the data format in that partition. This module is responsible for the execution process of the storage structure and provides atomic guarantee for the database system. Through the snapshots and logs in the database, the conversion of data partitions between different formats will become relatively simple.
\end{itemize}

\subsection{Workflow}
In the proposed system, there are two main workflows running through the five sub-modules of the storage engine selection module. Among them, the performance data collection and model learning workflow is used to establish the cost model for the evaluation step mentioned at the beginning of this section. The storage structure recommendation workflow is used to prune the workloads, and then generate, evaluate, and apply storage structures.
\subsubsection{Performance data collection and model learning}

Performance data is the record of the performance of the storage engine under various operations. The learning module uses the performance data to train and update the cost model. The performance data consists of two parts. Firstly, the system will perform a customized benchmark in the early stages of deployment, and the benchmark module will use the system behavior such as the elapsed time of each operation to construct the first batch of performance data. Since the benchmarks need to consider different data schemas, workloads, and storage engines, this process will take some time. If the system is deployed on servers with the same hardware configuration, the performance tests of the benchmark can be performed in parallel because the servers with the same hardware share the performance characteristic of the storage engine. The other part of the performance data is collected during system running. When users runs the database under real-world workloads, the system will sample the performance data. The sampled data is generated based on real scenarios, so it is more representative and used to enrich the performance data and update the cost models.

After collecting sufficient performance data in the benchmark, the benchmark module will not start until the hardware environment of the database changes. Whenever the database finds changes in the hardware environment, it is necessary to run the benchmark module and use the performance data collected by the benchmark to initialize the cost model. After initialization, the performance data will gradually increase, and the cost learning module will periodically check whether the performance data increases. If the performance data change, the cost learning module reruns the learning algorithm and updates the cost model.

\subsubsection{Storage Structure Recommendation}
The storage structure is defined as a combination of a storage engine and data layout. Fortunately, the choice of storage engines is relatively small, depending on the number of storage engines supported by the current database. However, the search space of data layouts is relatively large. Therefore, the system focuses more on the selection of the data layout. To evaluate the storage engine and data layout at the same time, the system will generate multiple choices of data layouts for the current workload of each data table. These data layouts represent local optimal solutions among all possible data layouts. They will be arbitrarily combined with possible storage engine and create all candidate storage structures. Then, the query engine of the database will generate a corresponding query plan based on the candidate storage structure and workload. The cost estimation module predicts the estimated cost of each candidate storage structure according to the query plan and selects the optimal storage structure among them.

Note that the optimal storage structure may not be applied immediately. The DBA can control the trigger conditions or the trigger time of the application process, such as a threshold of predicted performance improvement which DBA thinks is worthy to apply new storage structure, or a schedule that DBA thinks is suitable for storage changes.

\section{learned cost model}
To compare storage structures, a nature way is to establish a cost model for storage engines. Many methods for modeling the cost have been proposed, such as analysis method of addressing and scanning of disk \cite{c3} or learning method using query features \cite{c4}. In our system, since the storage structures under different workloads are required to be compared, the cost model have the following requirements:

\underline{Comparability}: The operating costs obtained under different storage engines should be comparable to each other. This means that some cost models based on query cardinality cannot be used.

\underline{Compatibility}: The cost model should be compatible with different storage engines. The purpose of the system is to use the richness and reliability of the existing query engine, so it cannot be limited to a specific query engine. The cost model should avoid using storage engine internal details.

\underline{Workload-oriented}: Compared with the goal of query optimization, the system concerns more about the total cost of the workload. Therefore, the cost model should represent the global performance during the execution of the workload.

This section first introduces the features used by the cost model in Section III-A. The training data will be collected based on these features. In order to collect training data, we propose a novel benchmark in Section III-B. Finally, in Section III-C, we introduced how to use features and training data to learn the cost model.

To achieve the comparability, our cost model selects the query elapsed time as the measure, because the query elapsed time directly reflects the delay of the system when responding to different requests and reducing the delay of the system is the primary task of system design. Besides, the query elapsed time can be proportional to the system's other resource usage, such as IO usage and memory usage. Another reason for choosing time as the cost measure is that many query features are theoretically proportional to query time empirically, such as the selectivity of query predicates. The linear relationship between these features and the target measure is easily captured by machine learning algorithms.

\subsection{Model Features}
Since the results of machine learning are strongly related to the quality of the input training data, it is necessary to provide high-quality training data. Since time is used as a measure of cost, we use performance data as the training data. In order to get good performance data, we need to define the features provided in the data. In this section, we first analyze how different factors affect database performance, and then show the features used in the proposed model.

The features of the model should select the factors that affect the query most, while following the requirements mentioned in the previous section. Many factors determine the running time of a query, which can be divided into the following categories.

\textbf{Hardware features:} These features describe the types and parameters of hardware resources that the database system can use during the runtime. The hardware resources of the server rarely change when running the database.

\textbf{Database configuration features:} Each database has different configuration knobs to tune for performance improvement.

\textbf{Workload features:} These features include table schema, metadata, and query attributes. They determine the logic cost of the query execution, such as the number of data blocks are read, the utilization of the cache, and so on.

\textbf{Runtime features:} These features are the specific usage of operating system components and database components during execution, such as the use of page cache and the hits of database buffer. Database-level features are related to the internal knowledge of the database, which are hardly to capture uniformly . System-level features use status of the operating system components, such as the utilization of the page cache and the read and write throughput of the disk.

In addition to the above features, unpredictable features are encountered during query execution, such as whether the data being queried is cached, whether the query is early returned by additional structure (such as Bloom filters), and whether hardware performance is jittery. These features, as well as the previously described features that require internal knowledge of the database, are not used in the cost model due to the difficulty of being captured by the proposed system.

In summary, the cost model selects workload features and system-level runtime features as the input of the machine learning model. These features are summarized in Table
\ref{table1}.

\begin{table}[!t]
\caption{Features of performance data}
\begin{center}
\begin{tabular}{|p{0.5in}|p{1.3in}|p{1.2in}|}
\hline
\textbf{Categories}&\textbf{Feature}&\textbf{Comment}\\
\hline
&Average Length of Rows&\\
\cline{2-3}
Table&Number of Key/Value Fields&\\
\cline{2-3}
Schema&Length of Key/Value Fields&\\
\cline{2-3}
Features&Number of Fixed-length / Variable-length Fields&\\
\cline{2-3}
&Length of Fixed-length / Variable-length Fields&\\
\hline
&Operation Type&\\
\cline{2-3}
\makecell[l]{Operation\\Features}&\makecell[l]{Operation Result Size}&\makecell[l]{Records returned by read\\operations or records\\inserted by update\\operations}\\
\cline{2-3}
&\makecell[l]{Operation Selectivity}&\makecell[l]{Selectivity of scan\\operations or affected key\\range of update operations}\\
\cline{2-3}
&\makecell[l]{Randomness of Data\\Insertion}&\makecell[l]{Normalized inversion\\number of key insertion\\sequence}\\
\hline
&Disk Read/Write Throughput&In a recent period\\
\cline{2-3}
System-level&Loaded Pages in Page Cache/ Total pages of table file&\\
\cline{2-3}
Runtime Features&Number of Files / Number of Level-1 Files / Number of Level-2 Files&Used for amplification analysis of LSM storage\\
\hline
\end{tabular}
\label{table1}
\end{center}
\end{table}

\subsubsection{workload features}

The system extracts the feature of the table schema and operations. When considering the table schema, we treat all fields in primary key as key fields, and others as value fields. To simplify the data types, the fields are divided into fixed-length fields and variable-length fields, because the database system cares about the storage form of the field rather than the way that the field is expressed. These features affect the number of actual data blocks accessing when reading data or indexes for a given query.

We disassemble the query into a series of data access operations. Data access operations are simpler than queries and easier in feature extraction. The features of data access operations include operation result size and operation selectivity. The reason for using selectivity is that we want to get the locality of the operation. The size of the result is directly related to the number of data block accesses, and the locality of the operation is related to whether the data block can be cached.

In addition, the randomness of the data upon insertion is also recorded in the features. The data inserted sequentially will be closer to the sorted array in the physical structure, so it will have higher efficiency in insertion and search. We denote the insertion sequence of keys as
\begin{math}
\pi
\end{math}. We use the inversion number of
\begin{math}
\pi
\end{math} to measure the randomness of data insertion and normalize it to [0,1]. If the data is inserted into the storage engine sequentially as the log, the randomness is 0, and if the data is inserted randomly, the randomness is 1. The formula for randomness is as follows. Note that order and reverse order are treated equally in randomness.
$$randomness(\pi)=\begin{cases}
    \frac{4\text{inv}(\pi)}{{|\pi|^2-|\pi|}} & \text{inv}(\pi)<\frac{{|\pi|^2-|\pi|}}{4} \\
    2-\frac{4\text{inv}(\pi)}{{|\pi|^2-|\pi|}} & \text{inv}(\pi)\geq\frac{{|\pi|^2-|\pi|}}{4}
\end{cases}$$

\subsubsection{runtime features}

For the runtime features, we mainly monitor the operating system page cache usage and the disk read/write throughput in the most recent period. Page caching is a general optimization method of the operating system. It uses free memory to cache recently opened files, reducing the number of disk access requests. In the database, different tables normally correspond to different data files. If a table is recently opened for data access, some data pages in the file corresponding to this table will enter the operating system page cache. The LRU cache strategy of the general database will also store the recently opened data blocks in the database cache. Therefore, monitoring the page cache can reflect the usage of each table in the cache of the storage engine, without caring about the cache design details.

Also, disk read and write throughput will also affect query performance. When the disk is busy, data access requests will be delayed due to waiting or scheduling. Therefore, recording the page cache and disk throughput of different storage engines can help analyze the read and write performance of different storage engines and the use of disk IO, thereby improving the accuracy of prediction.

Besides, the proposed system considers some runtime features that are not related to storage engine implementation details but related to storage engine classification. For example, for the LSM storage engine, the number of files in the first level, the number of files in the second level, and the total number of files are collected to analyze the read or write amplification phenomenon \cite{c6}.

\begin{figure}[!t]
\centering
\includegraphics[width=2.7in]{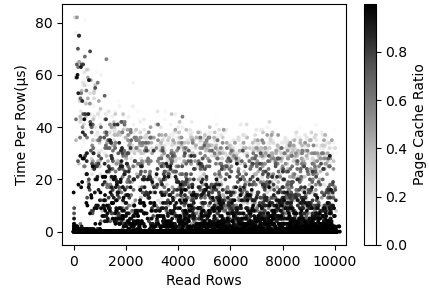}
\caption{Example of a figure caption.}
\label{figure2}
\end{figure}

In order to visually show the relationship between features and performance, we selected the two most obvious features of workload and runtime features for testing. The results are shown in
Fig. \ref{figure2}.

This graph is a scatter diagram of the operation time per row under the two characteristics of the number of scan lines and the operating system page cache utilization. It can be observed that with the increase in the number of scanning rows, the operation time per row is significantly reduced, and eventually tends to be stable. Under the same number of scanned rows, the higher the operating system page cache utilization rate is, the less the operation time per row is.

\subsection{Benchmark Design}
In the performance data collecting, we need to capture the features mentioned above, and ensure that the training data meets the requirements of machine learning in size and quality. Common benchmarks such as TPCH and YSCB focus on evaluating the performance of different databases under specific workloads, so they are not suitable for the proposed system. This paper proposes a brand-new benchmark method, which automatically generates a variety of workloads to obtain more comprehensive training data.The workflow of proposed benchmark in shown in Fig. 3.

\subsubsection{Data schema generation}
\begin{figure}[!t]
    \centering
    \includegraphics[width=3.4in]{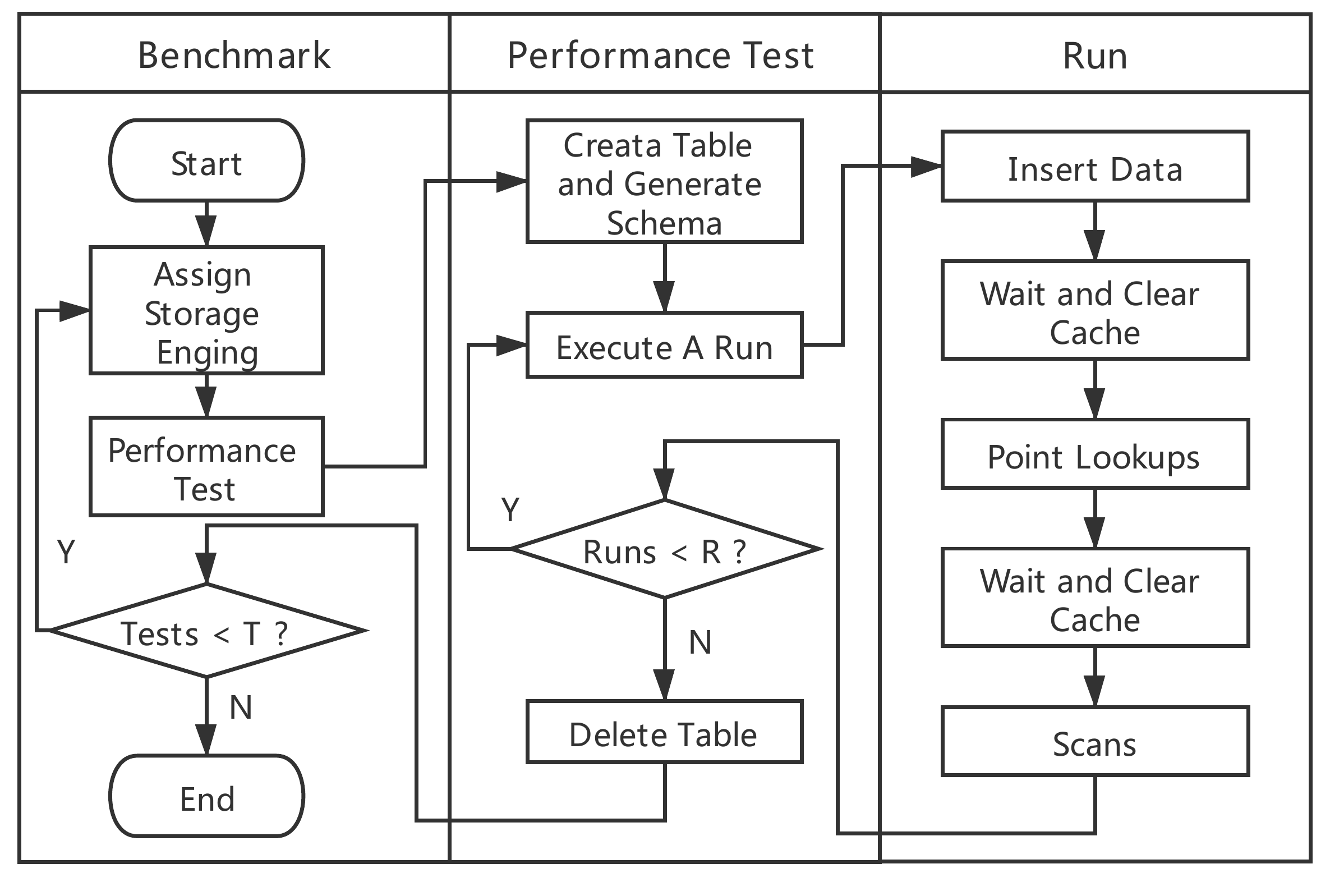}
    \caption{Workflow of the Benchmark}
\label{figure3}
\end{figure}

The data schema used in the proposed benchmark is randomly generated. The generation mainly considers the number, length, and type of fields. The data schema with the data generation pattern of each field will be assigned at the beginning of the performance test. The diversity of data schema is a guarantee for the sufficiency of training data. The numbers of fields are generated from a long-tail distribution within a fixed range, which refers to the actual scenario of the database application.

In the general data schema, each field also has a type attribute to determine the size and encoding method, and we used two types as fixed-length fields and variable-length fields. The size of the fixed-length field is designed to be relatively small bytes, considering that this kind of field is generally used to store numeric data. The variable-length field can take the length within a relatively large range as the average length of the field. In addition, the average length of the variable-length field is also taken from a long-tailed distribution, since most of the data fields store numbers or labels, while only a few data fields store long text.

Another reason of using long-tail distribution is that, in terms of database design, the database usually uses a fixed-size data block as the unit of operating data. When the row length is large, the difference of several bytes does not affect the number of rows stored in the data block. Therefore, longer rows are less influential than shorter rows for data blocks and have less weight in distribution.

When generating the data schema, the pattern of field content should also be designed, such as the data insertion order (log-type sequential insertion or transaction-type random insertion), the amount of information in the field (information entropy), etc. The design of these patterns can refer to real-world data.

\subsubsection{Workload of Performance Test}
After specifying the data schema, the benchmark module will insert and query the table
according to the workload. The workload of performance test should be
feasible and contain variable features. To reduce the complexity,
the workload is defined as a series of basic operations on the storage engine,
which are common access method\cite{c7}. If the storage engine supports special operations such as secondary index lookup, column aggregation, etc., additional cost models and training data are needed for engines that support these operations. For the convenience of discussions, this paper only introduces the design of common operations.

Although write operations are classified into single-row insertion and batch insertion, most storage engines that use disk as the storage media will buffer single-row insertion and convert it into batch updates using technology like write-ahead log. To specify write operation, we only need the parameter of write rows after data is already generated. Due to the small number of large-scale insertions in practice, the number of write rows should also be long-tail distribution.

The read operation is divided into point lookup and range scans. Since the database system uses data blocks as the operation unit, sequential scans will benefit from the locality, and the performance per row of point lookups will be significantly worse than that of range scans. When designing the read operation workload, it is necessary to pay attention to the impact of the cache. If the keys of two adjacent queries are also adjacent, the second query will read the data through the cache, thereby affecting the judgment of the time of a single read operation.

After the database runs for a while, some data may be cached in memory as "hot data", causing bias to the performance results. To avoid that most of the training data belong to the “hot data” query, the workload should be controlled. There are two ways to control the workload. One is to clear the operating system's page cache. The page cache is usually larger than the actual cache of the database, making data easily be “hot”. Clearing page cache can simulate the process of page cache replacement due to IO operations outside the target table. Another is to interrupt the workload and wait for a period to simulate the idleness of the database workload. These two operations will affect the characteristics of the page cache and the disk throughput in the operating system.

\subsection{Learning Model}

\begin{figure}[!t]
    \centerline{\includegraphics[width=1.6in]{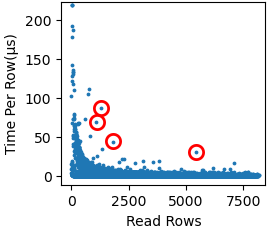} \includegraphics[width=1.6in]{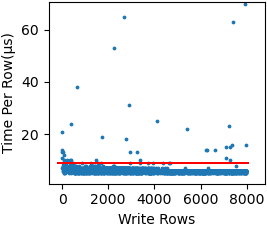}}
    \caption{Outliers under different operations}
\label{figure4}
\end{figure}

\begin{figure}[!t]
    \centering
    \includegraphics[width=3.4in]{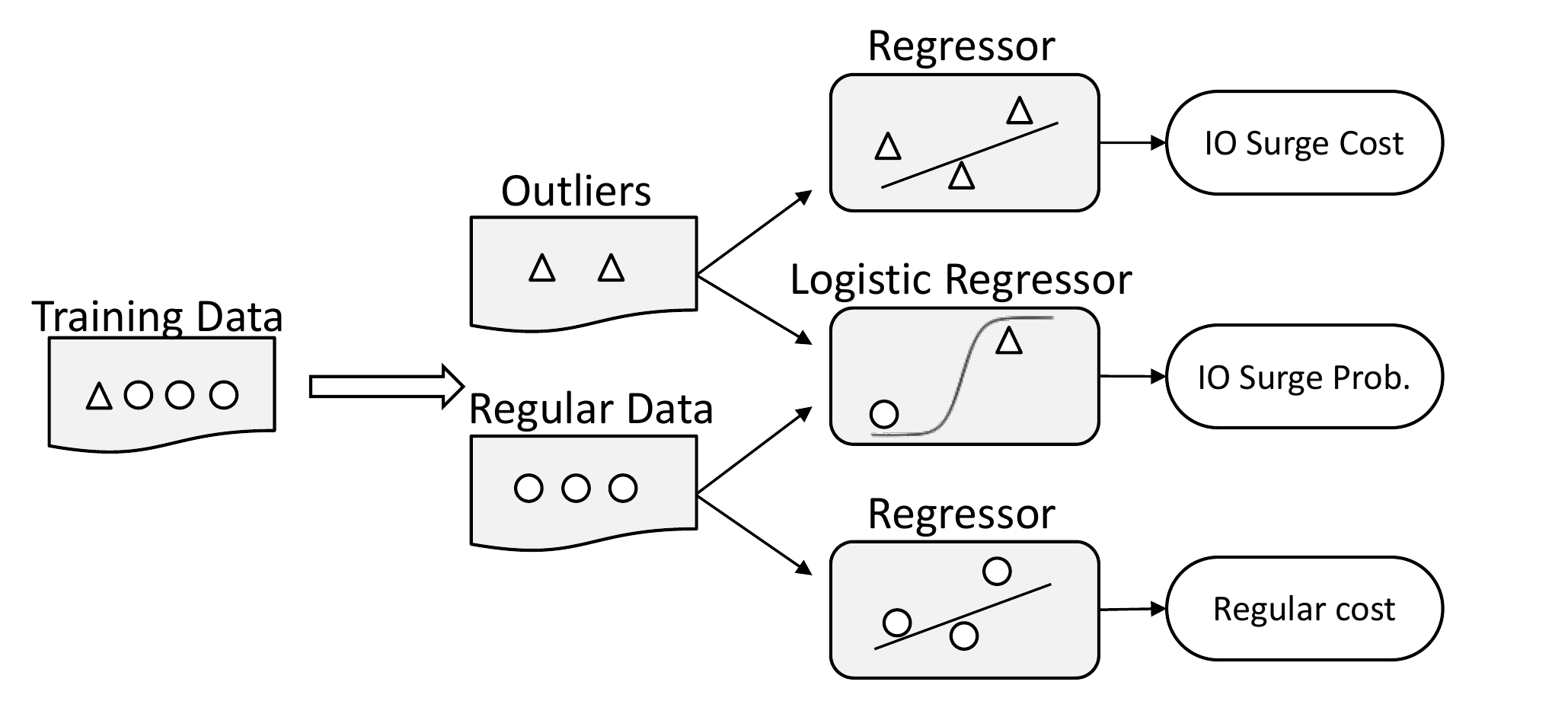}
    \caption{Cost model for writing}
\label{figure5}
\end{figure}

In this section, we use performance data to train the cost model. The training is divided into writing cost model training and reading cost model training. In order to obtain a more accurate cost model, we analyzed the collected performance data and designed the cost model according to the features of the performance data.

Performance data obtained by benchmark and system runtime are stored separately according to different classifications of storage engines and operations, because the cost model is designed to only predict certain operations on certain storage engine. They need to remove outliers before training. By observing the performance data, it is discovered that different operations have different performance features, as shown in
Fig. \ref{figure4}. For read performance, when the number of scan rows is low, the elapsed time per row increases significantly. The write performance is relatively stable because of the existence of the write buffer in the database. Therefore, for writing performance data, thresholds can be used to distinguish outliers, but for reading data, density clustering is more suitable for outlier discrimination.

However, the outliers in the write performance data do not mean that the point has no benefit on the storage engine model. Due to the existence of the write buffer, the data is accumulated after reaching a threshold, and then the write is performed. At this time, the emergence of IO increases the time of write operations, which is manifested in the surge of elapsed time per row. This surge has impact on the total performance. In our test of RocksDB, the query delay caused by the outliers in the write data accounts for about 5-10$\%$ of the total delay. Therefore, the writing cost model should take surge under consideration. We refer to the outliers in the write data as the IO surge point. The appearance of IO surge points is also considered to be related to the features in the training data.

When training writing cost models, we use multiple learners, as shown in
Fig. \ref{figure5}. The proposed writing cost model uses a logistics regression to learn the IO surge point to predict the probability of its occurrence under a given workload, because it is a parameter-free learner with high efficiency. Since the goal of the system is to estimate the cost of the workload, there should be massive write operations in the workload. According to the law of large numbers, the sum of the IO surge point expectation of write operations in the workload should be close to the actual number of IO surge events.

The writing cost model also establishes a regression model of costs for IO surge points and regular data points, to determine the elapsed time under the occurrence of IO surge events and regular write events respectively. These regression models select the XGBoost learner\cite{xgboost}, which is a trade-off between performance and prediction accuracy. Considering the feature of commercial servers, the servers may not have enough resources to run some machine learning algorithms, while XGBoost has lower computation requirements and has stronger accuracy than traditional learning methods. If the prediction of the elapsed time under regular write event is expressed as
$f_r(x)$, and the prediction of the elapsed time under the IO surge event is expressed as
$f_o(x)$, then the overall time prediction for the write workload is:
$$f_w (X)=\sum(f_o (x)p(x)+f_r (x))$$
The read operations are generally not performed after buffering multiple requests, so there is no need to consider modeling outliers. The reading cost model only uses the XGBoost learner for training.

After the query engine parses the SQL statement, the query plan tree composed of operators is generated, including operators related to data access and operators related to calculation. When predicting the workload, the system uses the query engine to generate the query plan tree for each query of the workload, and use the data access operators to predict the workload cost. For the storage structure using the same query plan tree, the order of the access operations prediction costs can represent the order of the query cost because the calculation operators are exactly the same. For the storage structure using different query plan trees, prediction errors may be caused by different calculation operators. In general, for data-intensive applications, the data access operator takes much longer time than the calculation operator. In this case, the error caused by the calculation operator has a relatively small influence on the cost order of the storage structure. Thus, for insurance, we set a threshold
$\epsilon$, e.g. 10
$\%$. If the best storage structure cannot exceed the predicted performance of the current storage structure by
$\epsilon$, this storage structure will not be applied in the database.

\section{Data Layout Recommendation}
In many studies of storage optimization \cite{c8}, data layout is one of the focuses of research. In our proposed automatic storage structure selection system, data layout is also considered to better utilize the storage engine. In this section, we first introduce the type of storage model in Section IV-A, and then propose a novel column-oriented data layout algorithm in Section IV-B.
\subsection{Storage Model and Workload}
The storage model refers to the logical arrangement of data in storage. There are three storage models in the current database design. The selection of storage model is closely related to the workload. N-ary Storage Model (NSM) for traditional databases and OLTP databases, Decomposition Storage (DSM) Model for OLAP databases, and Flexible Storage Model (FSM)\cite{c9} for hybrid workloads.

\begin{figure}[!t]
    \centerline{\includegraphics[width=3in]{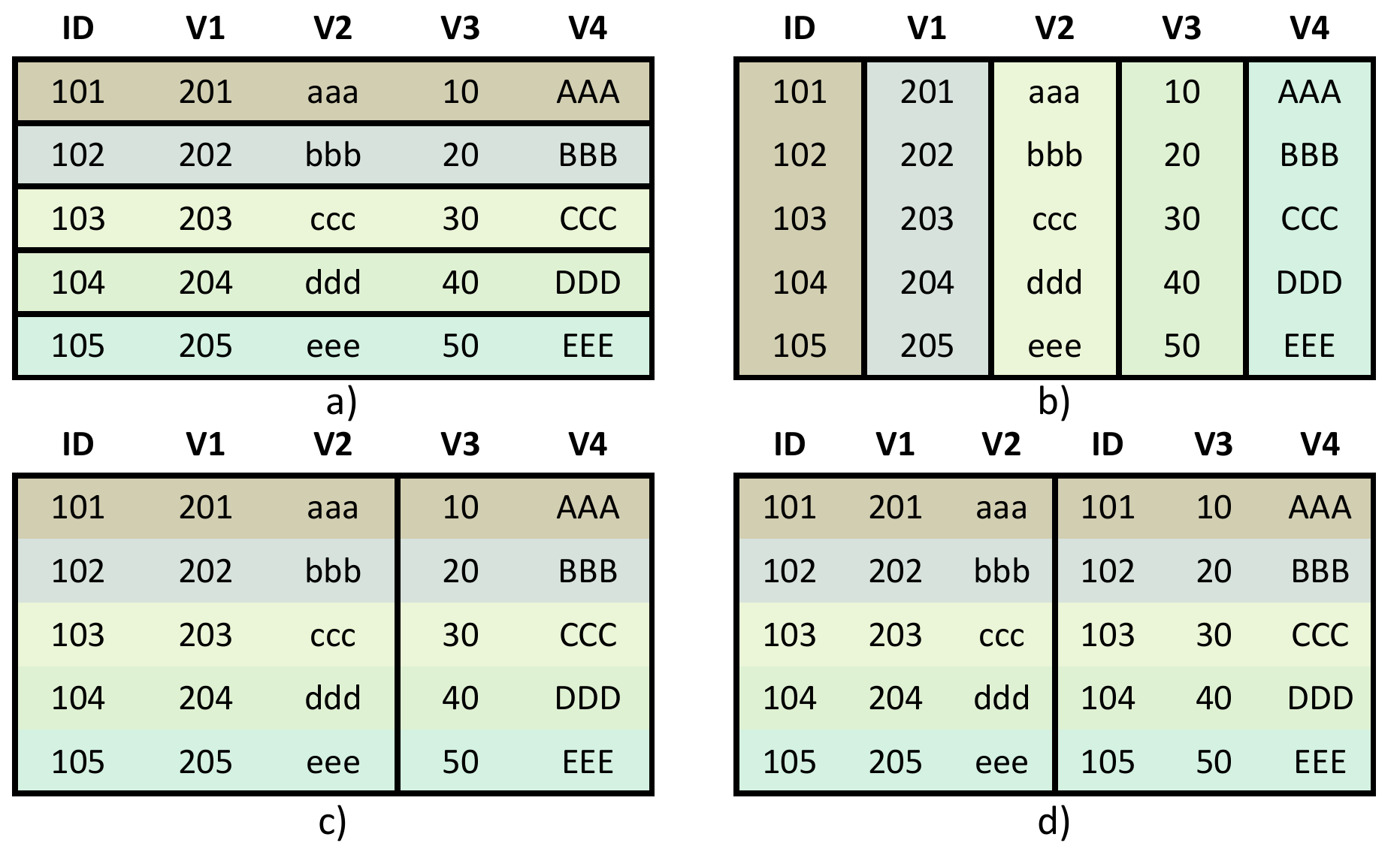}}
    \caption{Storage Models. a) NSM, b) DSM, c) FSM inside in-memory database, d) FSM inside disk storage.}
\label{figure6}
\end{figure}

In NSM, data is stored sequentially with rows as shown in
Fig. \ref{figure6}.a. Typical queries in OLTP like point lookup and data insertion are benefited from NSM because these queries have the least access to data blocks. In DSM, data is stored with cells that belong to the same column as shown in
Fig. \ref{figure6}.b. Cells in different columns are stored in separate files. In DSM, the cost of single row reads and writes becomes expensive, because these operations need to access multiple files for the whole cells, and the data access pattern is scattered which reduces the data locality. In contrast, column-wise aggregation operations no longer require access to irrelevant cells, but the cells on the target column. This feature greatly reduces the size of data to access. Besides, cells of the same type are stored together, which is suitable for compression and encoding. The workload of OLAP generally does not include point lookup and update of data, and analytical queries will only group and aggregate the columns of interest, so it is suitable for DSM storage models.

For a hybrid workload that contains both analytical and transactional queries, some studies combine the advantages of NSM and DSM to design the Flexible Storage Model (FSM). The intuitive idea is to divide the columns into multiple column groups as shown in
Fig. \ref{figure6}.c. Each column belongs to exactly one column group, which is easy for the in-memory database. The in-memory database only needs to expand the pointer in the index into a list of pointers, and each pointer in the list corresponds to one column group.

However, in the implementation of disk storage, another solution is used. In order to allow the storage engine to use the primary key to access the tuples in different column groups, each column group will hold the complete key fields as shown in
Fig. \ref{figure6}.d. Under this scheme, the increase in the number of column groups will lead to an increase in the redundancy of key fields. In FSM, it is necessary to balance the performance of transactional queries with the performance of representative analytical queries. For example, we found some important analytical queries in the workload, and hope to optimize these queries.  If we divide the table into many column groups, then these queries can access fewer columns to obtain the results. Since each column group will save a copy of keys, there will be a lot of redundancy in the storage, which reduce the performance of transactional queries.

\subsection{Data Layout Recommendation Algorithm}

The design of column groups is also called data layout. The data layout design is the main problem in the study of FSM. Only a good data layout can maximize the performance of the FSM database. The best data layout depends on the workload. To measure whether a data layout is optimal, the evaluation of the data layout cost under the target workload is needed. The solution space of the data layout problem is the
 \textit{Bell number}\footnote{\url{https://en.wikipedia.org/wiki/Bell_number}} of the column set. It is not feasible to traverse all possible data layouts to obtain the optimal result. Therefore, heuristic algorithms or cost-based algorithms are in demand, which search the solution using strategies mentioned in Section II.

\textbf{Pruning:} Use frequent item thresholds or clustering methods to reduce the huge number of queries in the workload to a few representative queries.

\textbf{Candidate generation:} For each representative query, the candidate data layout is column groups that make the representative query achieve the maximum performance. That is, all the columns accessed by the representative query are combined into a column group, and the unvisited columns are grouped separately. Data layouts obtained from different representative queries will be combined into a new data layout. For example, the columns accessed by the most important queries are grouped into the first column group. Among the remaining columns, the columns accessed by the second most important queries are combined into the second column group, and so on.

\textbf{Evaluation:} Evaluate the generated candidates of data layout using the cost model. The least cost candidate after evaluation is selected.

Among the data layouts generated by the above methods, the "primary" column group is always able to find its corresponding specific query (or representative query). This paper refers to these methods as query-oriented data layout recommendation algorithms.

In addition to the query-oriented data layout recommendation algorithm, this paper proposes a new perspective to solve the data layout problem, that is, the column-oriented data layout recommendation algorithm. The main difference between these algorithms is how candidates are generated. In the column-oriented algorithm, the candidate data layout is not regarded as a combination of optimal data layouts for a certain query. Instead, the data layout is regarded as the judgment of the similarity of the columns, and the similar columns are clustered to obtain the candidate data layouts.

\textbf{Example.} Consider the following query: T is a table with 5 columns (a, b, c, d, e). Q1 visits three columns (a, b, c) on T, and Q2 visits three columns (c, d, e) on T. Using a query-oriented algorithm, the candidate data layout obtained from Q1 and Q2 are either (a, b, c)(d, e), or (a, b)(c, d, e), because the result of Q1 and the result of Q2 are mutually exclusive and cannot be merged to generate new candidates.

In column-oriented algorithms, each column is represented as a vector consisting of query information. After pruning the query, representative queries are obtained. Each representative query contains a collection of columns that need to be accessed. The vector is composed of the representative query accessing the column. As in the previous example, the representative vector of the column is obtained as a = b = (1, 0), c = (1, 1) and d = e = (0, 1). The vectors of the columns represent the similarity between the columns during query access, and the columns that are close to each other are often accessed together by the queries in the workload. In this scenario, we get a new data layout (a, b)(c)(d, e), which is different from the previous algorithm and is beneficial to both queries. The column vector is expressed more specifically as follows:
$$c_{ij}=\sum_{q\in Q_j}{w_q I(c_i,q)}$$
$Q_j$ represents the
$j^{th}$ representative query, and
$c_ij$ represents the
$j^{th}$ dimension in column
$c_i$. When
$c_i$ is the column accessed by q,
$I(c_i,q)=1$, otherwise
$I(c_i,q)=0$. If the representative query is the central query of a series of queries, then these queries share one dimension of the column vector,
but each query can assign different values to the column vector according to the specific column accessed. In addition, the formula uses the cost
$w_q$ of the query to weight the vector, in which case columns accessed by costly queries are grouped preferentially. It should be pointed out that if the value of one dimension is composed of multiple queries, the accuracy of the vector to the column distance depends on the reasonableness of the clustering result. The reason is that it is impossible to distinguish which queries contributed the value in one dimension, and if the values are contributed by dissimilar queries, it leads to misjudgment of the distance between different columns.

In some algorithms, special methods are used to make the calculation result drift toward the recent queries. In the proposed algorithm, we use a time-dependent factor to achieve this purpose, where the weight of a query q is set to
$(1-\alpha)^s w_q$. In this formula, s is an attribute that indicates how old the query is. A larger value indicates that the query occurred earlier.
$\alpha$ represents the attenuation coefficient, and the larger the value is, the less attention is paid to old queries.

After obtaining the representative vectors of columns, we use the Euclidean distance to measure the similarity between different columns, and use clustering methods such as hierarchical clustering to generate column group candidates. The advantage of hierarchical clustering is that for the example above, if the cluster parameter is 3, the result is exactly the split of one of the column groups when the cluster parameter is 2. In this way, the maintenance cost caused by multiple column groups is weighed against the additional data read caused by fewer column families.

\begin{algorithm}[H]
\caption{DATA LAYOUT RECOMMENDATION}
\begin{algorithmic}[1]
\renewcommand{\algorithmicrequire}{\textbf{Input: Query set: Q, table: T}}
\renewcommand{\algorithmicensure}{\textbf{Output: Data layout candidates}}
\REQUIRE
\ENSURE \phantom{place holder}
\textit{Function DataLayoutRecommendation (Q, T)} :
\# get column feature vectors
 \FOR{$col_i$ in $T.col$}
 \STATE $col_i:=\boldsymbol{0}$
 \FOR{$q_j$ in $Q$}
 \IF{$col_i$ in $q_j.columns$}
 \STATE $col_{ij} := col_{ij} + q_j.cost$
 \ENDIF
 \ENDFOR
 \ENDFOR
 \STATE \# cluster columns
 \STATE $C:=\text{LevelCluster}(T.col)$
 \STATE $R := \emptyset$
 \FOR{$i$ in len($T.col$)}
 \STATE $R := R \cup C.\text{ExtractCluster}(i)$
 \ENDFOR
 \RETURN $R$
\end{algorithmic}
\end{algorithm}

Since the performance of the data layout is related to the implementation of specific storage engines, it is difficult to determine the number of column groups in advance. Therefore, in this algorithm, all hierarchical clustering results from the one group (that is, NSM) to the most groups i.e. DSM, will be given at the same time. The Cartesian product of the data layout result set and the storage engine set is computed as the candidate storage structure set, which is used for cost estimation. In the example presented in this section, assuming that Q1 has a greater weight than Q2, then there will be 4 data layouts, namely (a, b, c, d, e), (a, b, c) (d, e), (a, b)(c)(d, e), (a)(b)(c)(d)(e). If we have 3 storage engines to choose from, then we can get a total of 3×4=12 candidate storage structures. The final optimal storage structure is estimated and ranked according the cost model within these candidates.

\section{Experimental Evaluation}
\subsection{Experimental configuration}
The evaluation content is divided into three parts. The first part tests the storage engine cost model proposed in Section III. The second part tests the data layout recommendation algorithm proposed in Section IV. The third part evaluates the quality of storage structure recommended results.

Experiment run on a personal computer with an Intel Core i7 4790, a 3.6GHz×8 processor, 16GB of RAM, and a 7200 RPM hard drive.

The goal of the experiments is to optimize the storage structure of the LineItem table in the TPC-H benchmark\footnote{\url{http://www.tpc.org/tpch/}}. To run the experiment, we build the whole system. Our system uses two well-known storage engines, RocksDB with LSM tree as the storage model and WiredTiger with B+ tree as the major storage model. Both storage engines have very powerful performance and are used in many applications. WiredTiger is MongoDB's default storage engine and can be used independently. It supports columnar storage which will be considered as another storage engine in the experiment, meaning that the system has three different storage engines to choose.

\begin{table}[!t]
\caption{Columns accessed in different test queries}
\begin{center}
\begin{tabular}{|c|c|}
\hline
\textbf{Query}&\textbf{Columns Accessed}\\
\hline
Q3&K1, V2, V3, V7\\
\hline
Q5&K1, K3, V2, V3\\
\hline
WQ&K1-K4, V1-V12\\
\hline
RQ1&K1-K4, V1-V12\\
\hline
RQ2&K1-K3, V1-V3, V5-V8, V11 \\
\hline
\end{tabular}
\label{queries}
\end{center}
\end{table}

Since the purpose of the system is to recommend storage structures under hybrid workloads, we design workloads on LineItem table. It contains two analytical queries and three custom transactional queries. The analytical queries used the standard queries Q3 and Q5 in the TPC-H benchmark. Transactional queries consist of a single-row write query and two single-row read queries. We use K1-K4 to denote the first 4 columns (primary key) of the LineItem data schema, and use V1-V12 to denote the last 12 columns of the LineItem. The columns accessed by each query are shown in Table
\ref{queries}.

We designed four workloads based on the above queries, namely transactional workload, transactional mixed workload, analytical mixed workload, and analytical workload. The details of the workloads are shown in
Table \ref{diff_algo}.

\subsection{Storage Engine Cost Model}

The proposed system is based on the learning cost model, whose recommendation results depend on the accuracy of the cost model in estimating the workload. Also, we need to find out whether features of Section III are sufficient for cost prediction, and whether the learned cost model has enough generalization capabilities. To achieve this goal, we use the cost model which only learns the performance from generated data schema to make predictions on the performance of LineItem data.

We performed a series of data read and write operations according to the performance test mentioned in Section III.A. For the convenience of observation, we draw scatter plots with the accessed rows per operation as the x-axis and the elapsed per row as the y-axis, since the accessed rows per operation are the most important factor for the elapsed time. The results of the experiment are shown in Fig. 7. The green dots in the insets represent the actual time of the operation, and the blue dots represent the predicted time of the operation. The shapes of the blue dots and green dots in the insets are basically overlapped, indicating that the learned cost model can well describe the relationship between the elapsed time per row and the accessed rows in operations. Also, the shape of blue dots can fall within the shape of green dots, which means that although the model cannot get accurate time predictions, it can get average results. The above results demonstrate that data schemas and workloads generated by the benchmark algorithm can provide sufficient features to generalize the cost model. Although the cost model learned without using the internal knowledge of the storage engine cannot accurately predict the cost of a single operation, it can represent the average time of the workload.

\begin{figure}[!t]
    \centerline{\includegraphics[width=1.6in]{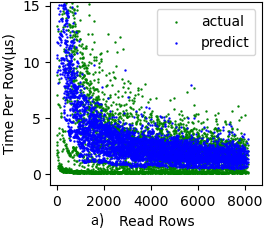} \includegraphics[width=1.6in]{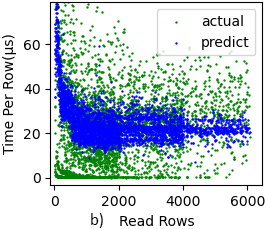}}
    \centerline{\includegraphics[width=1.6in]{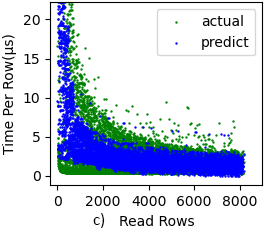} \includegraphics[width=1.6in]{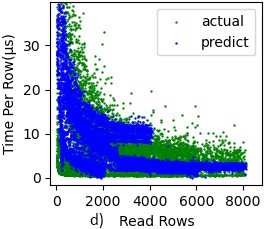}}
    \centerline{\includegraphics[width=1.6in]{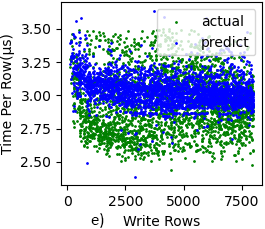} \includegraphics[width=1.6in]{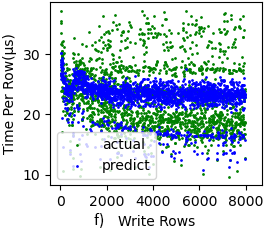}}
    \centerline{\includegraphics[width=1.6in]{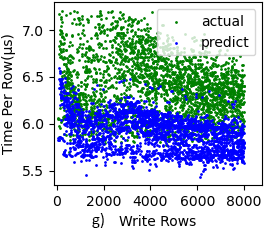} \includegraphics[width=1.6in]{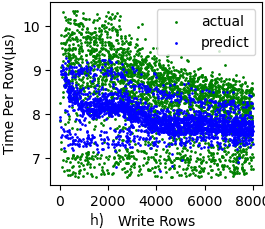}}
    \caption{Actual elapsed time vs. predicted time. Inset a), b), c), d) are using model of read operations while e), f), g), h) are using model of write operations. Inset a), b), e), f) are using WiredTiger as storage engine while c), d), g), h) are using RocksDB. Inset a), c), e), g) are using sequential insertion while b), d), f), h) are using random insertion.}
\label{figure7}
\end{figure}

We also show the mean absolute error for operation prediction and the relative error for workload prediction (using benchmark workload) in Table \ref{table4}. The relative error of the workload is defined as the relative error between the time to perform the benchmark test (excluding the waiting time) and the predicted time. The experimental results corroborate the above conclusion that the cost model may contain large errors when predicting a single operation. However, when making an overall prediction of the workload, the cost model can give no more than 10
$\%$ error.

\subsection{Data Layout Recommendation}
This section demonstrates whether the column-oriented algorithm can produce better data layout recommendation results as the query-oriented algorithm. We implemented a typical query-oriented data layout recommendation algorithm which was proposed in Peloton \cite{c9} to compare with our proposed algorithm. In the experiment, we run the algorithms with four types of workloads.

The comparison results are shown in
Table \ref{diff_algo}.
The vectors in the workload characteristics represent the number of corresponding queries executed. For example, the transactional workload includes 9M data insertions, 10k RQ1 queries, and 10k RQ2 queries. \textit{s} indicates the size of the table before the workload is executed. For each workload, the Peloton algorithm gives the optimal data layout design. Since our proposed algorithm is designed to be evaluated together with the storage engine, the results of the algorithm include the storage engines (omitted in the table) and only show the best two data layouts for convenience.

\begin{table}[!t]
\caption{Prediction error for different operations}
\begin{center}
\begin{tabular}{|p{1.2in}<{\centering}|p{1.0in}<{\centering}|p{0.8in}<{\centering}|}
\hline
\textbf{Operation of Storage}&\textbf{Mean absolute error for operation prediction ($\mu s$)}&\textbf{Relative error for workload prediction ($\%$)}\\
\hline
B+ / read / sequential&2.43&9.93\\
\hline
B+ / read / random&14.40&6.81\\
\hline
B+ / write / sequential&0.9&2.75\\
\hline
B+ / write / random&3.42&0.08\\
\hline
LSM / read / sequential&2.22&6.35\\
\hline
LSM / read / random&7.13&2.11\\
\hline
LSM / write / sequential&1.48&7.63\\
\hline
LSM / write / random&1.41&4.01\\
\hline
\end{tabular}
\label{table4}
\end{center}
\end{table}

\begin{table}[!t]
\caption{Data Layout Recommendation of Different Algorithm}
\begin{center}
\begin{tabular}{|p{0.8in}|p{1.1in}|p{1.1in}|}
\hline
\makecell[c]{\textbf{Workload}\\\textbf{characteristics}}&\makecell[c]{\textbf{Query-oriented}\\\textbf{Algorithm (Peloton)}}&\makecell[c]{\textbf{Column-oriented}\\\textbf{Algorithm (Proposed)}}\\
\hline
\makecell[l]{Transactional\\workload\\(9M,10k,10k,0,0)\\\textit{s}=0}&NSM&\makecell[l]{Best: NSM\\Second best:\\(V1-V3, V5- V8, V11),\\(V4, V9, V10, V12)}\\
\hline
\makecell[l]{Transactional\\mixed workload\\(3M,30k,30k,5,5)\\\textit{s}=9M}&NSM&\makecell[l]{Best: NSM\\Second best:\\(V1, V4-V6, V8-V12)\\(V2, V3, V7)}\\
\hline
\makecell[l]{Analytical\\mixed workload\\(0,10k,10k,15,15)\\\textit{s}=12M}&\makecell[l]{(V1, V4-V6,V8-V12),\\(V2, V3, V7)}&\makecell[l]{Best:\\(V1, V4-V6, V8-V12),\\(V2, V3, V7)\\Second best: NSM}\\
\hline
\makecell[l]{Analytical\\workload\\(0,0,0,30,30)\\\textit{s}=12M}&\makecell[l]{(V1, V4-V6,V8-V12),\\(V2, V3, V7)}&\makecell[l]{Best: DSM\\Second best:\\(V1, V4-V6, V8-V12),\\(V2, V3, V7)}\\
\hline
\end{tabular}
\label{diff_algo}
\end{center}
\end{table}

In Table \ref{diff_algo}, we find that the results obtained by the Peloton algorithm are included in the results of our proposed algorithm. This shows that the proposed column-oriented data layout algorithm can produce similar results as the query-oriented data layout algorithm. Noting that since the storage engines participate in the data layout cost evaluation, the proposed algorithm can obtain the DSM solution in the analytical workload which is not easy to obtain by query-oriented data layout algorithms. This also shows that the selection of data layout is closely related to the storage engine implementation, so it should not be designed independently.

\subsection{Storage Structure Selection}
\begin{figure}[!t]
    \centering
    \includegraphics[width=3.4in]{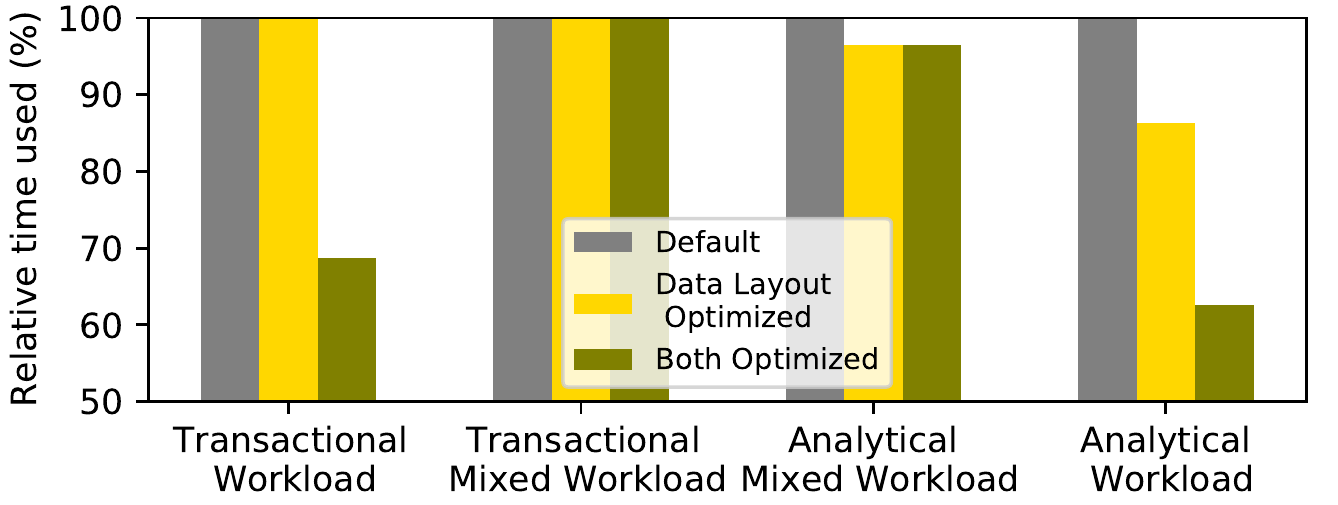}
    \caption{Performance comparison under different storage structure}
\label{figure8}
\end{figure}

Finally, we evaluate the storage structure recommended by the proposed system. We choose the RocksDB storage engine with NSM as the default storage structure design, and discuss the impact of different storage structures on the system performance under four workloads. We use the default design as a standard to regularize the time used by the recommended storage structure. We also observed the storage structure obtained with fixed storage engine to explore the importance of the storage engine selection in the storage structure. The results are shown in
Fig. \ref{figure8}.

Since the default storage structure uses NSM, the optimization of data layout only has a significant impact on workloads with a large number of analytical queries. That is, with the increase of analytical queries, the recommended data layout can execute the workload faster. The result also shows that if the storage engine and data layout are change at the same time, workload execution is faster than changing only the data layout. According to the results of Section V-C, the recommended data layouts are different under different storage engines, which means that the data layout and storage engine need to be decided together in storage structure recommendation. Only changing the data layout will not take full advantage of the changes in the storage structure.

We also observe that the optimized storage structure only achieves a greater degree of performance improvement in transactional and analytical workloads. The internal reason is that either WiredTiger and RocksDB is not designed for mixed workloads, and the upper limit of system performance is determined by the upper limit of these two storage engines. Besides, the proposed system only uses one storage engine in one partition, so it is natural to have poor performance in a mixed workload. The data partition still affected by the idea of “one size not fit all”.

\begin{figure}[!t]
    \centering
    \includegraphics[width=3in]{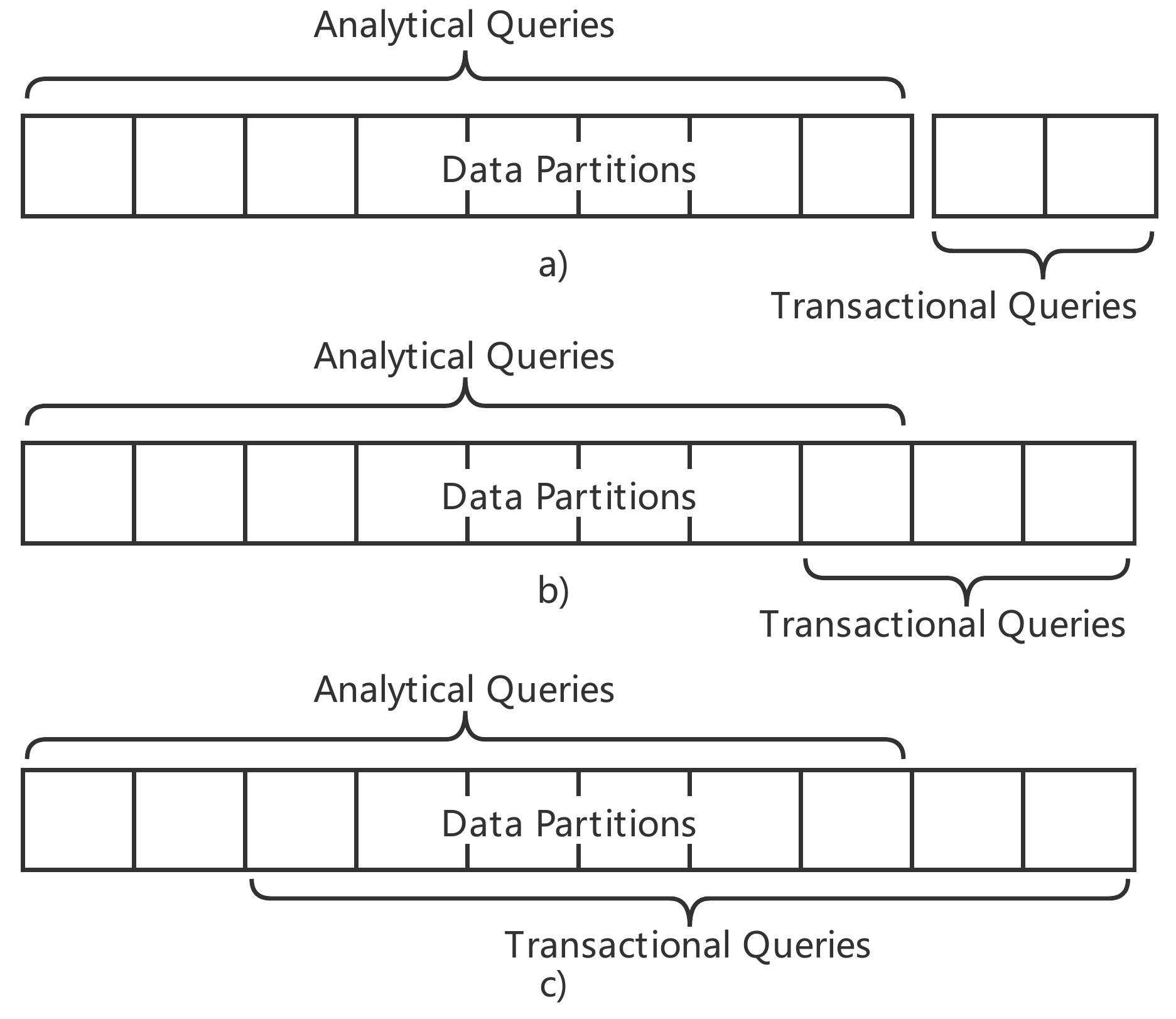}
    \caption{Different types of hybrid workloads}
\label{figure9}
\end{figure}

In hybrid workloads, although there are different types of queries, the data accessed by these queries may be different. In the scenario where the OLAP system and the OLTP system are used together, due to the independence of the systems, the data cannot be accessed from each other as shown in Fig. 9.a. But in the proposed system, there is allowed to be an intersection of the transactional queries and analytical queries on accessing data, as shown in Fig. 9.b. Data outside the intersection only needs to process a single type of query, which can benefit from the storage structure. This means that after horizontal partition of data is performed on such workloads, only a single type of workload exists in some data partitions which can benefit from the proposed system. For workloads with large intersections as shown in Fig. 9.c, the proposed system may not serve well due to the theoretical limitations of storage.

In either case, the storage structure selection algorithm proposed by our system can automatically select different storage structures for different partitions. So far, the proposed system has automatically completed all the work related to HTAP system storage design, including choosing the storage engine and data layout for every partition in database, and implement the conversion of the storage structure when workloads changed.

\section{conclusion}

This paper proposes a system that automatically selects the storage structure and uses the theory of self-driving databases to solve the above problems. We have integrated all the modules in one system, realized the mechanism of data conversion in different storage structures, and avoided the cross-system data ETL. The proposed system can dynamically analyze the workload and recommend the storage engine and data layout that are most suitable for current workload. The problem of workloads shift can also be solved under the proposed system.

We develop solutions to several key issues in the process of designing the system. We have established a storage engine cost model that does not depend on the internal knowledge of the storage engine. We propose benchmarking techniques to provide training data for the storage engine cost model, and use CPU-friendly learning algorithms for common commercial servers without special setting such as large memory or GPU. We regard the storage engine and data layout as a whole and propose a novel data layout recommendation algorithm to make full use of the features of different storage engines.

We build a prototype system for the experiment. We show that the proposed system can automatically select the appropriate storage engine and data layout according to the current workload. Limited by "One size not fit all", mixed workload performance under a single data partition cannot be significantly improved. However, compared to the HTAP system composed of separate databases, our system can perform automatic selection of storage structure, automatic conversion of data, and adaptation under changing workload. The system proposed in this paper is relatively preliminary, and there is still much room for improvement. If the space cost of the storage structure is considered together, the system will be able to draw a more appropriate recommendation result. In addition, the current cost model cannot predict the concurrent performance of queries well,
which can also be considered in future work.

\bibliographystyle{ieeetr}
\bibliography{refs.bib}

\end{document}